\documentclass[preprint,showpacs,preprintnumbers,amsmath,amssymb]{revtex4}
\usepackage{epsfig}
\usepackage{graphicx}
\usepackage{dcolumn}
\usepackage{bm}

\newcommand{\ord}{{\cal O}}
\def\beq{\begin{equation}}
\def\eeq{\end{equation}}
\def\eeqn{\end{equation}}
\newcommand\iden{\leavevmode\hbox{\small1\normalsize\kern-.33em1}}

\newcommand{\sq}{\sqrt{2}}
\newcommand{\bea} {\begin{eqnarray}}
\newcommand{\eea} {\end{eqnarray}}

\newcommand{\Gm}{\Gamma}
\newcommand{\sbt}{s_{\beta}}
\newcommand{\cbt}{c_{\beta}}
\newcommand{\tbt}{t_{\beta}}
\newcommand{\sbcb}{s_{\beta} c_\beta}

\let\jnfont=\rm
\def\NPB#1,{{\jnfont Nucl.\ Phys.\ B }{\bf #1},}
\def\PLB#1,{{\jnfont Phys.\ Lett.\ B }{\bf #1},}
\def\EPJC#1,{{\jnfont Eur.\ Phys.\ Jour.\ C }{\bf #1},}
\def\PRD#1,{{\jnfont Phys.\ Rev.\ D }{\bf #1},}
\def\PRL#1,{{\jnfont Phys.\ Rev.\ Lett.\ }{\bf #1},}
\def\MPLA#1,{{\jnfont Mod.\ Phys.\ Lett.\ A }{\bf #1},}
\def\JPG#1,{{\jnfont J.\ Phys.\ G }{\bf #1},}
\def\CTP#1,{{\jnfont Commun.\ Theor.\ Phys.\ }{\bf #1},}
\def\JHEP#1,{{\jnfont JHEP \ }{\bf #1},}
\def\NPPS#1,{{\jnfont Nucl.\ Phys.\ Proc.\ Suppl.\ }{\bf #1},}
\def\CPC#1,{{\jnfont Computl.\ Phys.\ Commun.\ }{\bf #1},}
\def\CPL#1,{{\jnfont Chin.\ Phys.\ Lett. }{\bf #1},}

\begin{document}

\title{\ \\[10mm] Higgs boson production in photon-photon collision at ILC:\\
                  a comparative study in different little Higgs models }

\author{Lei Wang$^1$, Fuqiang Xu$^{2}$, Jin Min Yang$^{3}$}

\affiliation{
$^1$ Department of Physics, Yantai University, Yantai 264005, PR China\\
$^2$ Department of Physics, National Tsinghua University, Hsinchu, Taiwan 300, RO China\\
$^3$ Key Laboratory of Frontiers in Theoretical Physics,\\
     Institute of Theoretical Physics, Academia Sinica,
             Beijing 100190, PR China}


\begin{abstract}
We study the process $\gamma\gamma\to h \to b\bar{b}$ at ILC as a
probe of different little Higgs models, including the simplest
little Higgs model (SLH), the littlest Higgs model (LH), and two
types of littlest Higgs models with T-parity (LHT-I, LHT-II).
Compared with the Standard Model (SM) prediction, the production
rate is found to be sizably altered in these little Higgs models
and, more interestingly, different models give different
predictions. We find that the production rate can be possibly
enhanced only in the LHT-II for some part of the parameter space,
while in all other cases the rate is suppressed. The suppression can
be $10\%$ in the LH and as much as $60\%$ in both the SLH and the
LHT-I/LHT-II. The severe suppression in the SLH happens for a large
$\tan\beta$ and a small $m_h$, in which the new decay mode $h\to
\eta\eta$ ($\eta$ is a light pseudo-scalar) is dominant; while for
the LHT-I/LHT-II the large suppression occurs when $f$ and $m_h$ are
both small so that the new decay mode $h\to A_H A_H$ is dominant.
Therefore, the precision measurement of such a production process at
the ILC will allow for a test of these models and even distinguish
between different scenarios.
\end{abstract}


\maketitle

\section{Introduction}
Little Higgs theory \cite{LH} has been proposed as an interesting
solution to the hierarchy problem. So far various realizations of
the little Higgs symmetry structure have been proposed
\cite{otherlh,lst,sst}, which can be categorized generally into
two classes \cite{smoking}. One class use the product group,
represented by the littlest Higgs model (LH) \cite{lst}, in which
the SM $SU(2)_L$ gauge group is from the diagonal breaking of two
(or more) gauge groups. The other class use the simple group,
represented by the simplest little Higgs model (SLH) \cite{sst},
in which a single larger gauge group is broken down to the SM
$SU(2)_L$. However, due to the tree-level mixing of heavy and
light mass eigenstates, the electroweak precision tests can give
strong constraints on this model \cite{cstrnotparity,f4.5,f5.6},
which would require raising the mass scale of new particles to be
much higher than TeV and thus reintroduce the fine-tuning in the
Higgs potential. To tackle this problem, a discrete symmetry
called T-parity is proposed \cite{tparity}, which forbids those
tree-level contributions to the electroweak observables. For the
LH, there are two different versions of implementing T-parity in
the top quark Yukawa interaction. In the pioneer version of this
model (hereafter called LHT-I) \cite{lhti}, the T-parity is simply
implemented by adding the T-parity images for the original top
quark interaction to make the Lagrangian T-invariant. A
characteristic prediction of this model is a T-even top partner
which cancels the Higgs mass quadratic divergence contributed by
the top quark. An alternative implementation of T-parity has been
proposed (hereafter called LHT-II) \cite{lhtii}, where all new
particles including the heavy top partner responsible for
cancelling the SM one-loop quadratic divergence are odd under
T-parity. The implementation of T-parity in the SLH model has also
been tried \cite{slmtparity}.

These little Higgs models mainly alter the property of the Higgs
boson and hence the hints of these models can be unravelled from
various Higgs boson processes. The Higgs decay and main production
channels at the LHC have been studied in the SLH
\cite{higgsslhlh,slhvdefine,rrbbslh,higgsslh}, the LH
\cite{higgsslhlh,higgslh,hrrhan}, and the LHT-I
\cite{lhtiyuan,higgslhti} and LHT-II \cite{higgslhtii}. While the
LHC is widely regarded as a discovery machine for Higgs boson and
also could possibly allow for a measurement of decay partial
widths at $10\%-30\%$ level \cite{LHCprecis}, a precision
measurement of Higgs property can be only achieved at the proposed
International Linear Collider (ILC). With the ILC, the Higgs
nature can be scrutinized through the production in photon-photon
collision, where the photon beam can be obtained by backscattering
a laser light with high energy $e^{\pm}$ beam. Such an option of
photon-photon collision can possibly measure the rates of Higgs
productions with a precision of a few percent. Especially, for
$\gamma\gamma \to h \to b\bar{b}$ process, the production rate
could be measured at about $2\%$ for a light Higgs boson
\cite{ilcprecis,ilcprecis130-150}.

Such a process $\gamma\gamma \to h \to b\bar{b}$ is a sensitive
probe for new physics because both the loop-induced $h\gamma\gamma$
coupling and the $h b\bar{b}$ coupling are sensitive to new physics.
Considering the sizable alteration of Higgs couplings in various
little Higgs models, we in this work study the process
$\gamma\gamma\to h \to b\bar{b}$ as a probe of different little
Higgs models, including the SLH, LH, LHT-I and LHT-II. Note that
this process has been studied in the LH \cite{rrbblh} and also in
the SLH \cite{rrbbslh}. In our study we give a comprehensive and
comparative analysis for all these models. In addition, since a
recent study of $Z$ leptonic decay gave a new stronger bound on the
parameter $f$ in the SLH \cite{f5.6}, we will consider such a new
bound in our calculation for the SLH.

This work is organized as follows. In Sec. II we recapitulate the
models. In Sec. III we calculate the rate of $\gamma\gamma \to h \to
b\bar{b}$ in these models. Finally, we give our conclusion in Sec.
IV.

\section{little Higgs models}
\subsection{Simplest little Higgs model}
The SLH \cite{sst} model is based on
$[SU(3) \times U(1)_X]^2$ global symmetry. The gauge symmetry $SU(3)
\times U(1)_X$ is broken down to the SM electroweak gauge group by
two copies of scalar fields $\Phi_1$ and $\Phi_2$, which are
triplets under the $SU(3)$ with aligned VEVs $f_1$ and $f_2$. The
uneaten five pseudo-Goldstone bosons can be parameterized as
\beq
\Phi_{1}= e^{ i\; t_\beta \Theta } \left(\begin{array}{c} 0 \\
0 \\ f_1
\end{array}\right)\;,\;\;\;\;
\Phi_{2}= e^{- i\; t_\beta \Theta} \left(\begin{array}{c} 0 \\  0 \\
f_2
\end{array}\right)\;,
\label{paramet}
\end{equation}
where
\begin{equation}
   \Theta = \frac{1}{f} \left[
        \left( \begin{array}{cc}
        \begin{array}{cc} 0 & 0 \\ 0 & 0 \end{array}
            & H \\
        H^{\dagger} & 0 \end{array} \right)
        + \frac{\eta}{\sqrt{2}}
        \left( \begin{array}{ccr}
        1 & 0 & 0 \\
        0 & 1 & 0 \\
        0 & 0 & 1 \end{array} \right) \right],
\end{equation}
with $f=\sqrt{f_1^2+f_2^2}$ and $t_\beta\equiv tan\beta= f_2 / f_1$.
Under the SM $SU(2)_L$ gauge group, $\eta$ is a singlet CP-odd
scalar, while $H$ transforms as a doublet and can be identified as
the SM Higgs doublet. The other five Goldstones are eaten by new
gauge bosons $Z'$, $W'_{0,\bar{0}}$, and $W'^{\pm}$, which obtain
masses proportional to $f$: \beq m^2_{W^{'+}} = \frac{g^2}{2}f^2,~~
m^2_{W^{'0}} = \frac{g^2}{2}f^2,~~
m^2_{Z'}=g^2f^2\frac{2}{3-tan^2\theta_W}, \eeq with $\theta_W$ being
the electroweak mixing angle.

The gauged $SU(3)$ symmetry promotes the SM fermion doublets into
$SU(3)$ triplets. There are two possible gauge charge assignments
for the fermions: the 'universal' embedding and the 'anomaly-free'
embedding.  Since the first choice is not favored by the electroweak
precision data \cite{sst}, we focus on the second way of embedding.
The top, strange, and down quarks have heavy partner quarks $T$,
$S$, and $D$, respectively. The mixing between light quarks and
heavy partners can be parameterized by \beq x_\lambda^t\equiv
{\lambda_1^t \over \lambda_2^t},\ \ \ x_\lambda^{d}\equiv
{\lambda_1^{d} \over \lambda_2^{d}},\ \ \ x_\lambda^{s}\equiv
{\lambda_1^{s} \over \lambda_2^{s}}. \eeq To leading order, the
heavy partners have masses proportional to $f$: \beq
m_Q=\sqrt{(\lambda_1^q c_\beta)^2+(\lambda_2^q s_\beta)^2} f, \eeq
where $Q=T,~D,~S$; $q=t,~d,~s$;
$c_\beta=\frac{f_1}{\sqrt{f_1^2+f_2^2}}$,
$s_\beta=\frac{f_2}{\sqrt{f_1^2+f_2^2}}$; $\lambda_1^q$ and
$\lambda_2^q$ are two dimensionless couplings of $q$-quark Yukawa
sector.

The Yukawa and gauge interactions break the global symmetry and then
provide a potential for the Higgs boson. However, the
Coleman-Weinberg potential alone is not sufficient since the
generated Higgs mass is too heavy and the new CP-odd scalar $\eta$
is massless. Therefore, one can introduce a tree-level $\mu$ term
which can partially cancel the Higgs mass
\beq
-\mu^2
(\Phi^\dagger_1 \Phi_2 + h.c.) = - 2 \mu^2 f^2 \sbt\cbt \cos\left(
\frac{\eta}{\sq \sbt\cbt f} \right)
 \cos \left(
 \frac{\sqrt{H^\dagger H}}{f \cbt\sbt}
\right).
\end{equation}
Then the scalar potential becomes
\beq \label{eq:VCW}
V = - m^2 H^\dagger H + \lambda (H^\dagger H)^2
 - \frac{1}{2} m_\eta^2 \eta^2 +\lambda' H^\dagger H \eta^2 + \cdots,
\end{equation}
where
\beq \label{eq:msq:lambda}
m^2 = m_0^2 - \frac{\mu^2}{\sbcb}, \quad
\lambda =\lambda_0 - \frac{\mu^2}{12\sbt^3 \cbt^3f^2}, \quad
\lambda' = - \frac{\mu^2}{4 f^2 \sbt^3 \cbt^3},
\end{equation}
with $m_0$ and $\lambda_0$ being respectively the one-loop
contributions to the Higgs boson mass and the quartic couplings from
the contributions of fermion loops and gauge boson loops \cite{sst}.
The Higgs VEV, the Higgs boson mass and the mass of $\eta$ are given
by
\beq \label{eq:vsq:mH:meta}
v^2 = \frac{ m^2}{\lambda} , \quad
m_h^2 = 2 m^2 , \quad m_\eta^2 = \frac{\mu^2}{\sbcb} \cos\left(
\frac{v}{\sqrt{2} f \sbcb} \right).
\end{equation}
The Coleman-Weinberg potential involves the following parameters:
\beq \label{para}
f,~ x_\lambda^t,~ t_\beta,~\mu,~m_\eta,~m_h,v.
\end{equation}
Due to the modification of the observed $W$-boson mass, $v$ is
defined as \cite{slhvdefine}
\beq \label{eq:v}
v \simeq v_{SM}
\left[ 1+ \frac{v_{SM}^2}{12 f^2}\frac{\tbt^4-\tbt^2+1}{\tbt^2} -
\frac{v_{SM}^4}{180 f^4}\frac{\tbt^8-\tbt^6+\tbt^4-\tbt^2+1}{\tbt^4}
\right],
\end{equation}
where $v_{SM}=246$ GeV is the SM Higgs VEV. Assuming that there are
no large direct contributions to the potential from physics at the
cutoff, we can determine other parameters in Eq. (\ref{para}) from
$f$, $t_\beta$ and $m_h$ with the definition of $v$ in Eq. (\ref{eq:v}).

\subsection{Littlest Higgs model}
The LH model \cite{lst,lhhantao} is based on a non-linear $\sigma$ model
in the coset space of $SU(5)/SO(5)$ with additional local gauge
symmetry $[SU(2) \otimes U(1)]^2$. A VEV of an $SU(5)$ symmetric
tensor field breaks the $SU(5)$ to $SO(5)$ at the scale $f$ with
\beq
\Sigma_0 \,=\, \left(\begin{array}{ccc}
0& 0& \iden\\
0& 1& 0\\
\iden& 0& 0\\
\end{array}\right).
\eeq
The non-linear sigma fields are then parameterized by the
Goldstone fluctuations as
\begin{equation}
\Sigma \simeq \Sigma_0 + \frac{2 i}{f} \left( \begin{array}{ccccc}
\phi^{\dagger} & \frac{H^{\dagger}}{\sqrt{2}} &
{\mathbf{0}}_{2\times 2} \\
\frac{H^{*}}{\sqrt{2}} & 0 & \frac{H}{\sqrt{2}} \\
{\mathbf{0}}_{2\times 2} & \frac{H^{T}}{\sqrt{2}} & \phi
\end{array} \right) + {\cal O}\left(\frac{1}{f^2}\right),
\end{equation}
where $H$ is a doublet and $\phi$ is a triplet under the unbroken
$SU(2)_L$. The other four Goldstones are eaten by new gauge bosons
$W^{\pm}_H$, $Z_H$, and $A_H$, which get masses of order $f$:
\begin{equation}
  m_{Z_H} = m_{W_H} = \frac{gf}{2sc},
  \qquad
  m_{A_H} = \frac{g^{\prime}f}{2\sqrt{5}s^{\prime}c^{\prime}},
  \label{eq:heavygaugemasses}
\end{equation}
with $c$, $s$, $c'$ and $s'$ being the mixing parameters in the gauge
boson sector given by
\begin{eqnarray}
\label{eq:Lst:gaugemixing}
    c \equiv \cos\theta  = \frac{g_1}{\sqrt{g_1^2+g_2^2}}\,, &\qquad&
    s \equiv \sin\theta  = \frac{g_2}{\sqrt{g_1^2+g_2^2}}\,, \nonumber \\
    c^{\prime} \equiv \cos\theta^{\prime}
        = \frac{g_1^{\prime}}{\sqrt{g_1^{\prime 2} + g_2^{\prime 2}}}\,, &\qquad&
    s^{\prime} \equiv \sin\theta^{\prime}
        = \frac{g_2^{\prime}}{\sqrt{g_1^{\prime 2} + g_2^{\prime 2}}}\,.
\end{eqnarray}
Here $g_j$ and $g'_j$ are the $SU(2)_j$ and $U(1)_j$ ($j=1,2$) gauge
coupling constants, respectively.

The top quark loops, gauge boson loops and scalar particles loops
can generate the Higgs potential, which trigger electroweak symmetry
breaking. The heavy bosons can further mix with light bosons,
leading the masses of heavy and light gauge bosons corrected at
$\ord{(\frac{v^2}{f^2})}$. The components $\Phi^{++}$, $\Phi^+$,
$\Phi^0$ and $\Phi^P$ (neutral pseudo-scalar) of the triplet $\phi$
get a mass
\beq
m_{\Phi}=\frac{\sqrt{2}m_h}{\sqrt{1-x^2}}\frac{f}{v},
\eeq
where $x$ is a free parameter of the Higgs sector proportional to the triplet
VEV $v'$ and defined as $x = \frac{4fv'}{v^2}$ with $v$ being the LH Higgs VEV
given by \cite{hrrhan}
\beq
v \simeq v_{SM}[1-\frac{v^2_{SM}}{f^2}(-\frac{5}{24}+\frac{1}{8}x^2)].
\eeq
In the fermion sector, there is an extra top quark partner
$T$-quark, which cancels the Higgs mass one-loop quadratic divergence
contributed by the top quark. The mixing between $t$ and $T$ can be
parameterized by
\beq
r=\frac{\lambda_1}{\lambda_2},~~
c_t=\frac{1}{\sqrt{r^2+1}},~~ s_t=\frac{r}{\sqrt{1+r^2}},
\eeq
where $\lambda_1$ and $\lambda_2$ are two dimensionless couplings of top
quark Yukawa sector. Together with $f$, the parameters can control
the $T$-quark mass
\beq
m_T=\frac{m_tf}{s_t c_t v}.
\eeq

\subsection{Littlest Higgs models with T-parity}
In the LHT-I \cite{lhti,lhtiyuan,lhtihubisz}, the T-parity is
simply implemented by adding the T-parity images for the original
top quark interaction to make the Lagrangian T-invariant. A
characteristic prediction of this model is a T-even top partner
which cancels the Higgs mass quadratic divergence contributed by
the top quark. Inspired by the way that the top quadratic
divergence is cancelled in the SLH, Ref. \cite{lhtii} takes an
alternative implementation of T-parity in LHT-II, where all new
particles including the heavy top partner responsible for
cancelling the SM one-loop quadratic divergence are odd under
T-parity. Thus, Higgs couplings with top quark and partners in the
two models have sizable difference. Besides, for each SM quark
(lepton), a copy of mirror quark (lepton) with T-odd quantum
number is added in order to preserve the T-parity. The Higgs
couplings with the down-type T-odd fermions are absent, and the
couplings with the up-type T-odd fermions are different in LHT-I
and LHT-II. For the above reasons, LHT-I and LHT-II can give
distinct predictions for production rates of single Higgs,
Higgs-pair, as well as a Higgs boson associated with a pair of top
and anti-top quarks at LHC \cite{higgslhtii}.

For the SM down-type quarks (leptons), the Higgs couplings have two
different cases \cite{lhtiyuan}
\begin{eqnarray}
\frac{g_{hd\bar{d}}}{g_{hd\bar{d}}^{\rm SM}}
&\simeq&1-\frac{1}{4}\frac{v_{SM}^2}{f^2}+\frac{7}{32}
\frac{v_{SM}^4}{f^4} ~~~~{\rm for~Case~A}, \label{Higgs-downA} \nonumber\\
&\simeq&1-\frac{5}{4}\frac{v_{SM}^2}{f^2}-\frac{17}{32}
  \frac{v_{SM}^4}{f^4} ~~~~{\rm for~Case~B}.\nonumber
\label{eq15}
\end{eqnarray}
The relation of down-type quark couplings also applies to the lepton
couplings.

The LHT-I and LHT-II have the same kinetic term of $\Sigma$ field
where the T-parity can be naturally implemented by setting $g_1=g_2$
and $g'_1=g'_2$. Under T-parity, the SM bosons are T-even and the
new bosons are T-odd. Therefore, the coupling of $H^{\dag}\phi H$ is
forbidden, leading the triplet VEV $v'=0$. In both LHT-I and LHT-II, the
Higgs VEV $v$ is modified as \cite{lhtiyuan,higgslhtii}
\beq
v\simeq v_{SM}(1+\frac{1}{12}\frac{v^2_{SM}}{f^2}).
\eeq

\section{The process $\gamma\gamma \to h \to b\bar{b}$ in
         little Higgs models}
\subsection{Calculations}
We consider a photon-photon collision at the ILC with the photon
beams obtained by Compton backscattering of lasers from the
$e^{\pm}$ beams. The cross section $\sigma(\gamma\gamma\to h)$ at
the ILC is obtained by folding the cross section
$\hat{\sigma}_{\gamma\gamma\to h}(\hat{s})$ with the photon
luminosity
 \beq
 \sigma(\gamma\gamma \to h)=\int_{0}^1d\tau\int_{\tau}^1
 \frac{dx}{x}f_{\gamma /e}(x)f_{\gamma /e}(\tau /x)
 \hat{\sigma}_{\gamma\gamma\to h}(\hat{s}),
 \eeq
where $f_{\gamma/e}(x)$ is the energy spectrum of the back-scattered photon
\cite{photonfunc}. The cross section $\hat\sigma$ is given by
\beq
\hat{\sigma}_{\gamma\gamma\to h}(\hat{s})
 =\frac{8\pi^2}{m_h}\Gamma(h\to\gamma\gamma)\delta(\hat{s}-m^2_h),
\eeq
where $\hat{s}=\tau s$ with $\sqrt{s}$ being the center-of-mass
energy of the ILC. The rate of $\gamma\gamma\to h\to b\bar{b}$
can be approximately obtained by
$\sigma(\gamma\gamma \to h) \times BR(h\to b\bar{b})$.
So, we need to calculate both the production cross section
$\sigma(\gamma\gamma \to h)$ and the decay widths.

Now we discuss the Higgs decays in little Higgs models. For the
tree-level decays $h\to f \bar{f}$ (SM fermion pair), $WW$ and $ZZ$,
the little Higgs models give the correction via the corresponding
modified couplings
\beq
\Gamma(h \to XX)= \Gamma(h \to
XX)_{SM}(g_{hXX}/g_{hXX}^{SM})^2,
\end{equation}
where $XX$ denotes $WW$, $ZZ$ or fermion pairs, $\Gamma(h \to
XX)_{SM}$ is the SM decay width, and $g_{hXX}$ and $g_{hXX}^{SM}$ are
the couplings of $hXX$ in the little Higgs models and SM,
respectively.

The loop-induced decay $h \to gg$ will be also important for a low
Higgs mass. The effective coupling of  $hgg$ is presented in
Appendix A.  In the SM, the main contributions are from the top quark
loop, and the little Higgs models give the corrections via the
modified couplings $ht\bar{t}$. In addition, the decay width of
$h\to gg$ can be also corrected by the loops of heavy partner quark
$T,~D$ and $S$ in SLH  ($T$ quark in LH) (new T-even and T-odd
quarks in LHT-I and LHT-II).

For the decay $h \to \gamma\gamma$, the main contributions are from
the top quark loop and $W$-boson loop in the SM. The little Higgs
models give the corrections via the modified couplings $ht\bar{t}$
and $hWW$. In these models the new quarks which contribute to the
decay $h\to gg$ also contribute to the decay $h\to \gamma\gamma$. In
addition to the contributions from fermion loops, the decay width of
$h\to \gamma\gamma$ can be also corrected by the loops of $W'$ in
the SLH ($W_H$, $\Phi^+$, $\Phi^{++}$ in the LH, LHT-I and LHT-II).
The effective coupling of  $h\gamma\gamma$ can be found in
Appendix A.

In addition to the SM decay modes, the Higgs boson in the SLH, LHT-I and
LHT-II has some new important decay modes which are kinematically allowed
in the parameter space. In the SLH, the new decay modes
are $h\to \eta\eta$ and $h \to Z\eta$, whose partial widths are
given by
\bea \label{eq:Gamma:new}
\Gm(h \to \eta\eta) &=&
 \frac{{\lambda'}^2}{8\pi}\frac{v^2}{m_h} \sqrt{1-x_\eta},\nonumber\\
\Gamma( h \to Z \eta) &=& \frac{m_h^3}{32 \pi f^2}
  \left( t_\beta - \frac{1}{t_\beta} \right)^2 \,
  \lambda^{3/2} \left(1, \frac{m_Z^2}{m_h^2}, \frac{m_\eta^2}{m_h^2}
 \right ),
\eea
where $x_\eta =4m_\eta^2/m_h^2$ and $\lambda (1,x,y) = (1-x-y)^2 - 4 xy$.
In the LHT-I and LHT-II, the new decay mode is $h\to
A_H A_H$, whose partial width is
\bea
\Gamma(h \to A_{H} A_{H}) & = & \frac{g_{hA_H A_H}^{2} m_h^3}{128 \pi m_{A_H}^4}
  \sqrt{1-x_{A_H}}\left(1-x_{A_H}+\frac{3}{4}x_{A_H}^2\right),
\eea
where $x_{A_H}=4m_{A_H}^2/m_{h}^2$, and $g_{hA_H A_H}$ is the
coupling constants of $hA_HA_H$. Note that the breaking scale $f$ in
LHT-I may be as low as 500 GeV \cite{flht-i}, and the constraint in
LHT-II is expected to be even weaker \cite{lhtii}. Therefore, for a
lower value of $f$, the lightest T-odd particle $A_H$ may have a
light mass, $m_{A_H}< \frac{m_h}{2}$, leading to the decay
$h\to A_H A_H$. However,  in the LH the electroweak precision data requires
$f$ larger than a few TeV \cite{cstrnotparity} and thus
the decay $h\to A_HA_H$ is kinematically forbidden.

In our calculations, the SM input parameters involved are taken from
\cite{pdg}. For the SM decay channels, the relevant higher order QCD
and electroweak corrections are considered using the code Hdecay
\cite{hdecay}. In the SLH, the new free parameters are $f,
~t_\beta,~ x_\lambda^{d}~(m_D)$ and $x_\lambda^{s}~(m_S)$. As shown
above, the parameters $x_\lambda^t,~\mu,~m_\eta$ can be determined
by $f$, $t_\beta$, $m_h$ and $v$. The small mass of the $d$ ($s$)
quark requires one of the couplings $\lambda^{d}_1$ and
$\lambda^{d}_2$ ($\lambda^{s}_1$ and $\lambda^{s}_2$) to be very
small, so there is almost no mixing between the SM down-type quarks
and their heavy partners.  We assume $\lambda^{d}_1$
($\lambda^{s}_1$) is small, and take $x_\lambda^{d}=1.1\times
10^{-4}$ ($x_\lambda^{s}=2.1\times 10^{-3}$), which can make the
masses of $D$ and $S$ in the range of 1-2 TeV with other parameters
fixed as in our calculation. In fact, our results show that the
contributions from $d$ and $D$ ($s$ and $S$) are very small compared
with the effects from $t$ and $T$. The electroweak precision data
can give a strong constraint on the scale $f$. Ref.\cite{sst} shows
that the LEP-II data requires $f>2$ TeV. In addition, the
contributions to the electroweak precision data can be suppressed by
large $t_\beta$. Ref. \cite{f4.5} gives a lower bound of $f>4.5$ TeV
from the oblique parameter $S$, while a recent study of $Z$ leptonic
decay gives a stronger bound of $f>5.6$ TeV \cite{f5.6}. Considering
the above bounds, in our numerical calculation we will take several
values of $t_\beta$ for $f=2$ TeV, $f=4$ TeV and $f=5.6$ TeV.

In the LH model, the new free parameters involved are $f,~c_t~(r),~c,~c'$ and
$x$, where
\beq
0<c_t<1,~~~0<c<1,~~~0<c'<1,~~~0<x<1.
\eeq
Taking $f=1$ TeV, $f=2$ TeV and $f=4$ TeV, we will scan over these parameters
in the above ranges and show the scatter plots.
Note that the widths $\Gamma(h\to t\bar{t})$, $\Gamma(h\to gg)$ and
$\Gamma(h\to \gamma\gamma)$ involve the parameter $c_t$ which can
control Higgs couplings with $t$, $T$ and $m_T$. For a light
Higgs boson, the decay mode $h\to t\bar{t}$ is kinematically
forbidden. For the decay $\Gamma(h\to gg)$ and $\Gamma(h\to
\gamma\gamma)$, the $c_t$ dependence of top-quark loop can cancel
that of T-quark loop to a large extent \cite{rrbblh}. Therefore,
the rate $\sigma(\gamma\gamma\to h)\times BR(h\to b\bar{b})$ is
not sensitive to $c_t$ for a light Higgs boson.

In LHT-I and LHT-II, the parameters $c$, $c'$ and $x$ are fixed as
\beq
c=c'=\frac{1}{\sqrt{2}},~~~x=0.
\eeq
The heavy T-even and T-odd
quarks only have large contributions to the decay widths of
$h\to gg$ and $h\to \gamma\gamma$, which are not sensitive to the actual
values of their masses as long as they are much larger than half of
the Higgs boson mass \cite{hrrhan}. Similar to the LH model, the result is
not sensitive to $c_t$ in LHT-I and LHT-II. Taking
$c_t=1/\sqrt{2}$ ($\lambda_1=\lambda_2$) can simplify the
top quark Yukawa sector in the LHT-II \cite{lhtii,higgslhtii}, and
this choice is also favored by the electroweak precision data
\cite{flht-i}. Therefore, in our numerical calculations we take
$c_t=1/\sqrt{2}$.

\subsection{Discussions}
\begin{figure}[tb]
 \epsfig{file=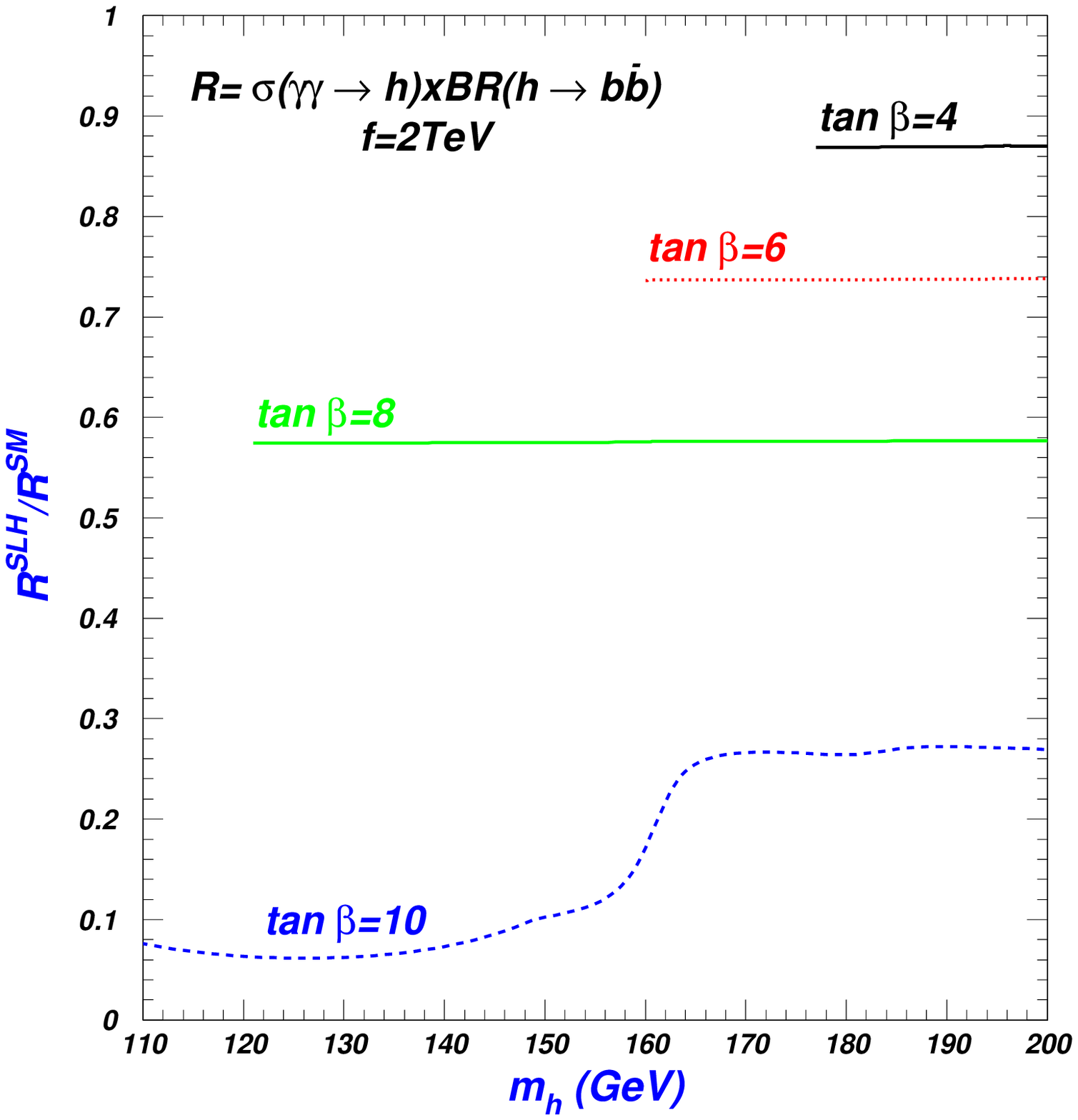,height=5.7cm}
 \epsfig{file=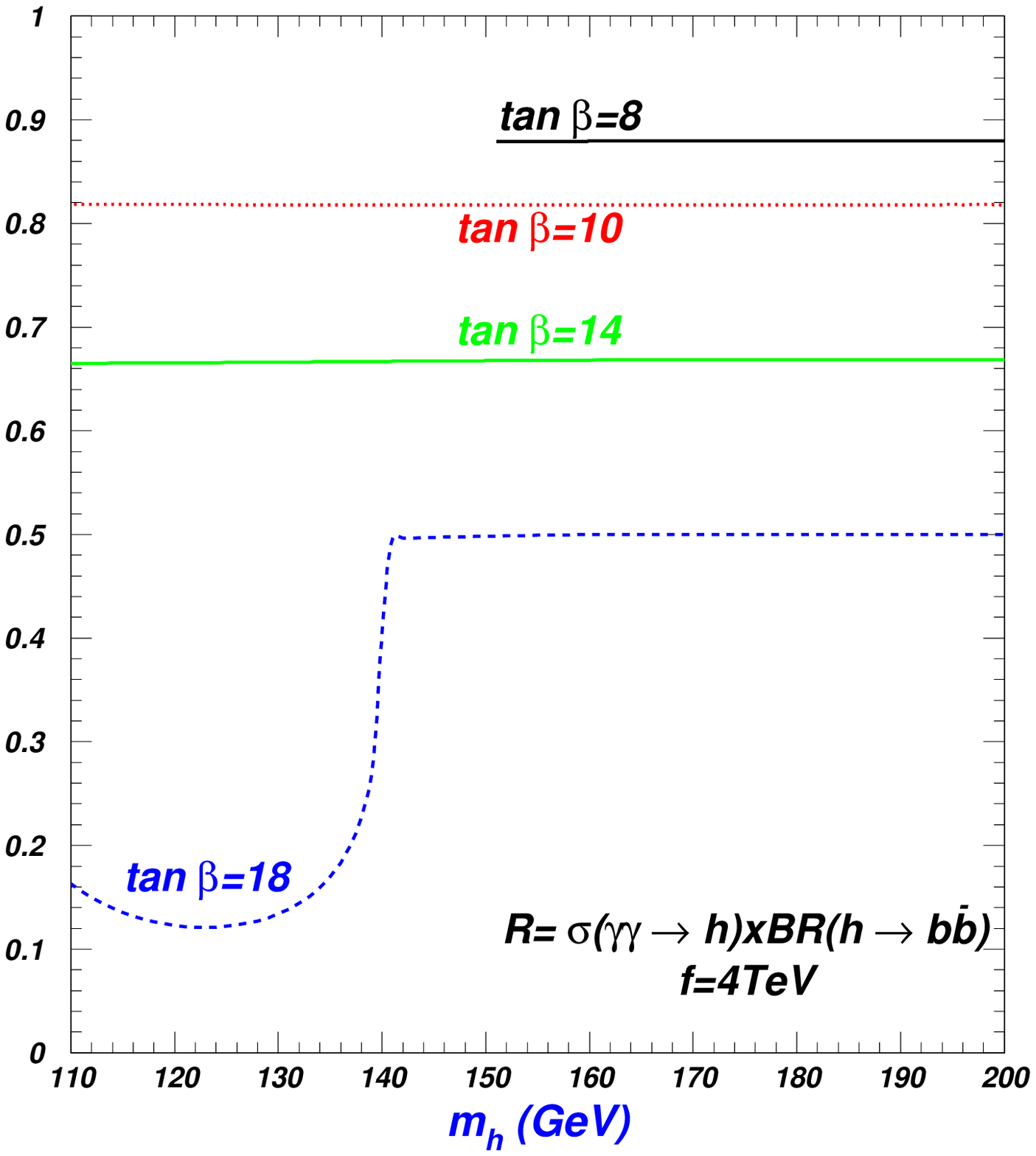,height=5.7cm}
 \epsfig{file=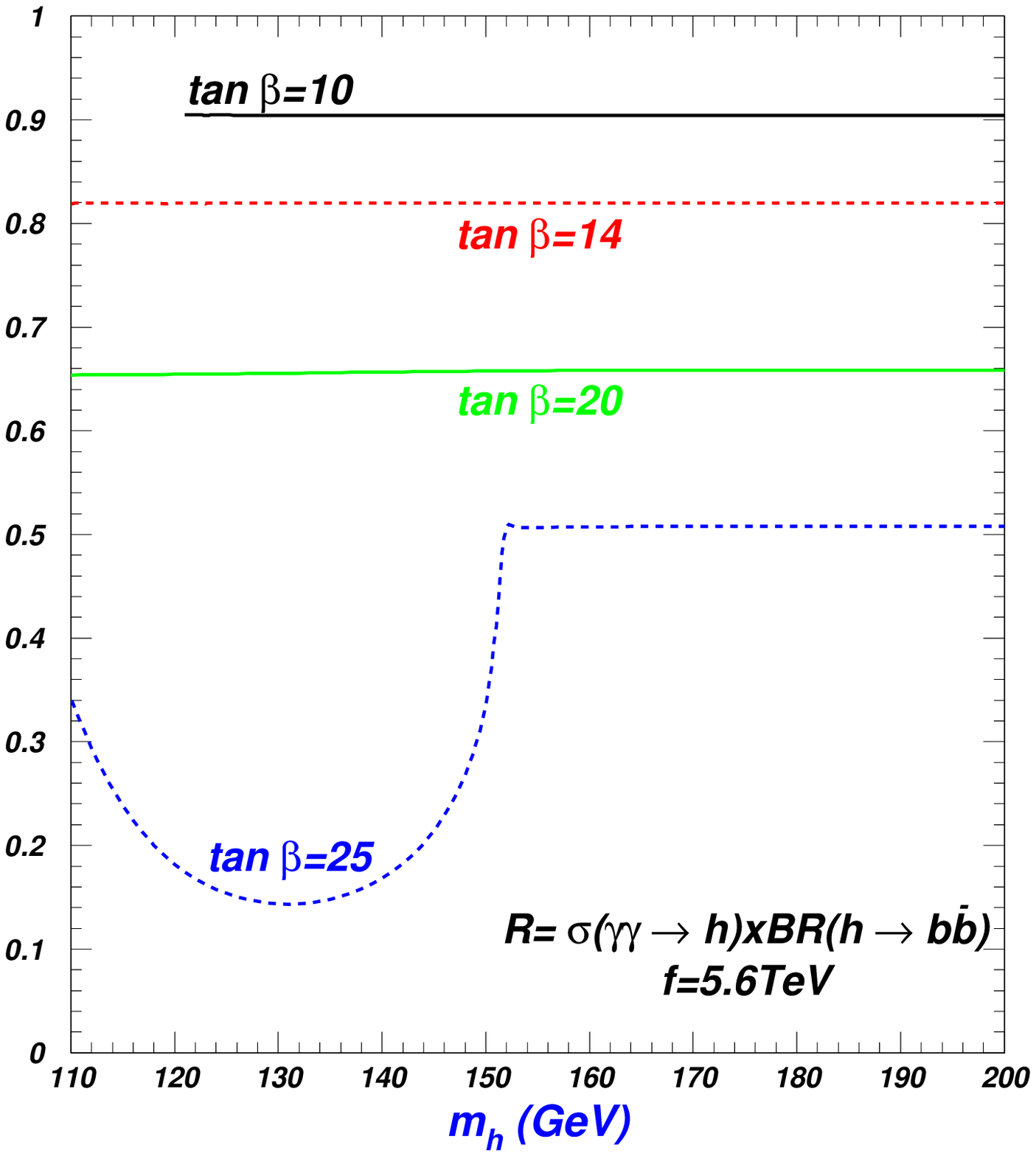,height=5.7cm}
\vspace{-0.5cm}
\caption{The rate $\sigma(\gamma\gamma\to
h)\times BR(h\to b\bar{b})$ normalized to the SM prediction in the SLH
model. The incomplete lines for small values of $\tan\beta$ show the
lower bounds of Higgs mass. }
\label{figslh}
\end{figure}

\begin{figure}[tb]
 \epsfig{file=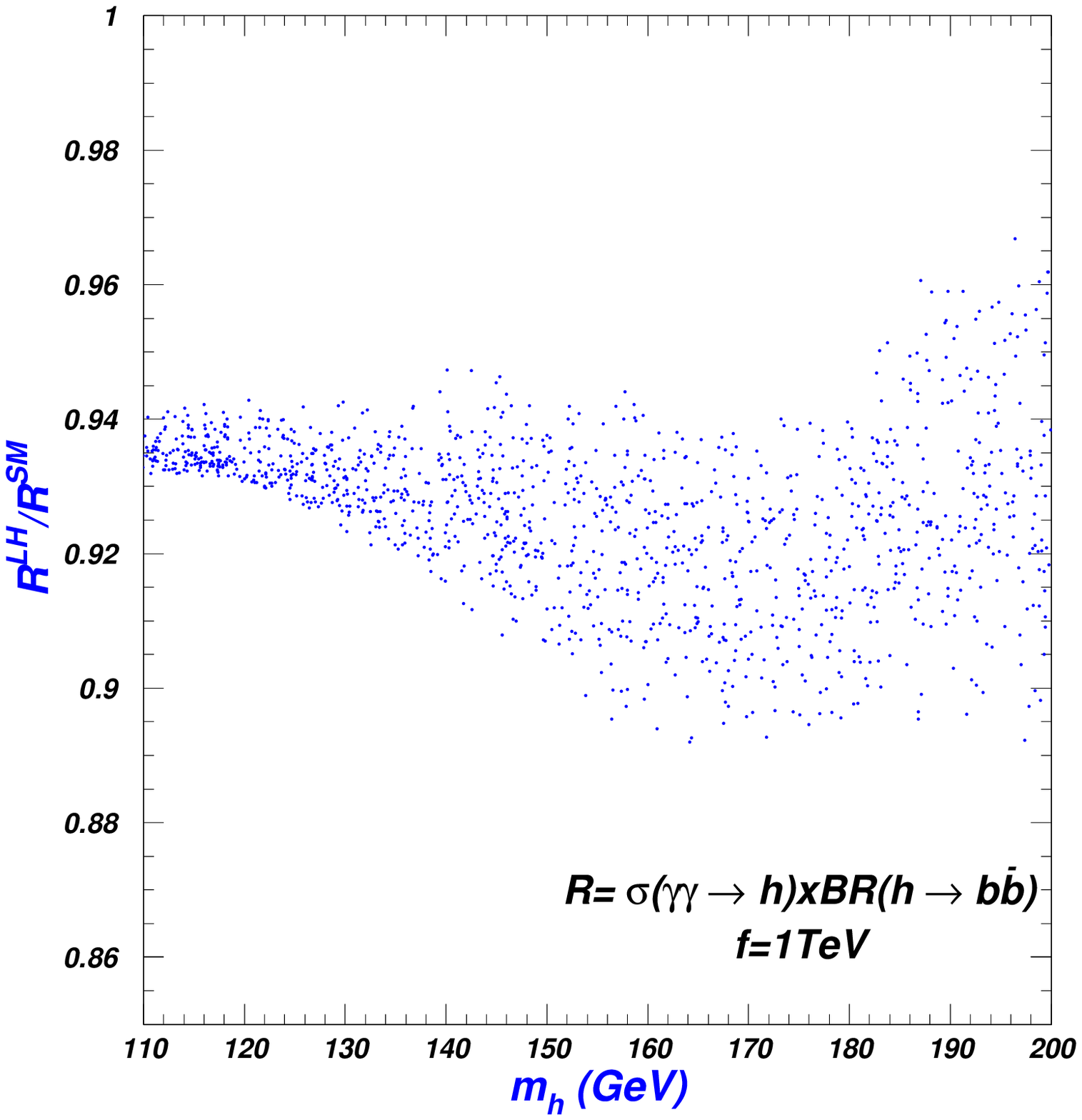,height=5.7cm}
 \epsfig{file=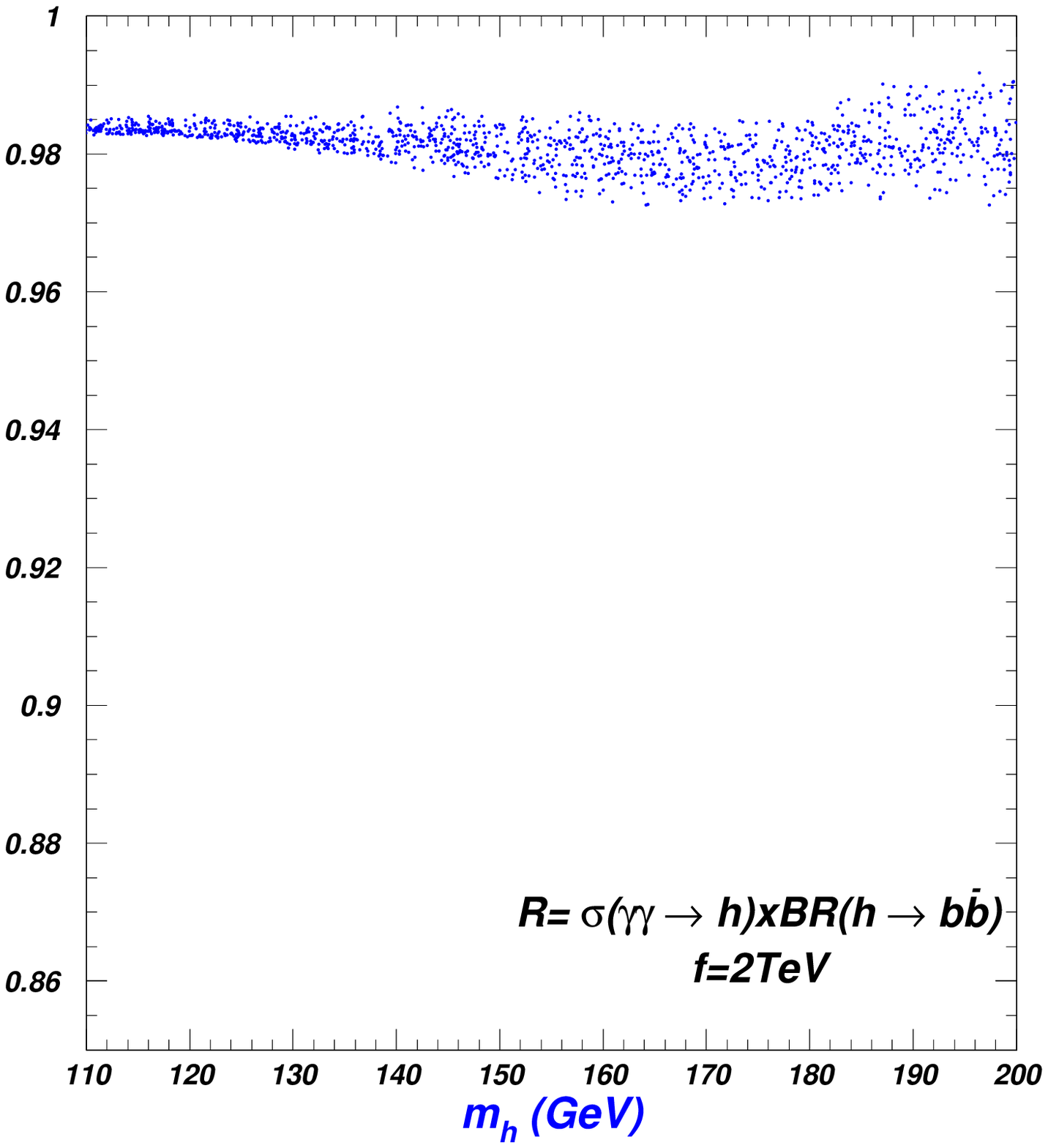,height=5.7cm}
 \epsfig{file=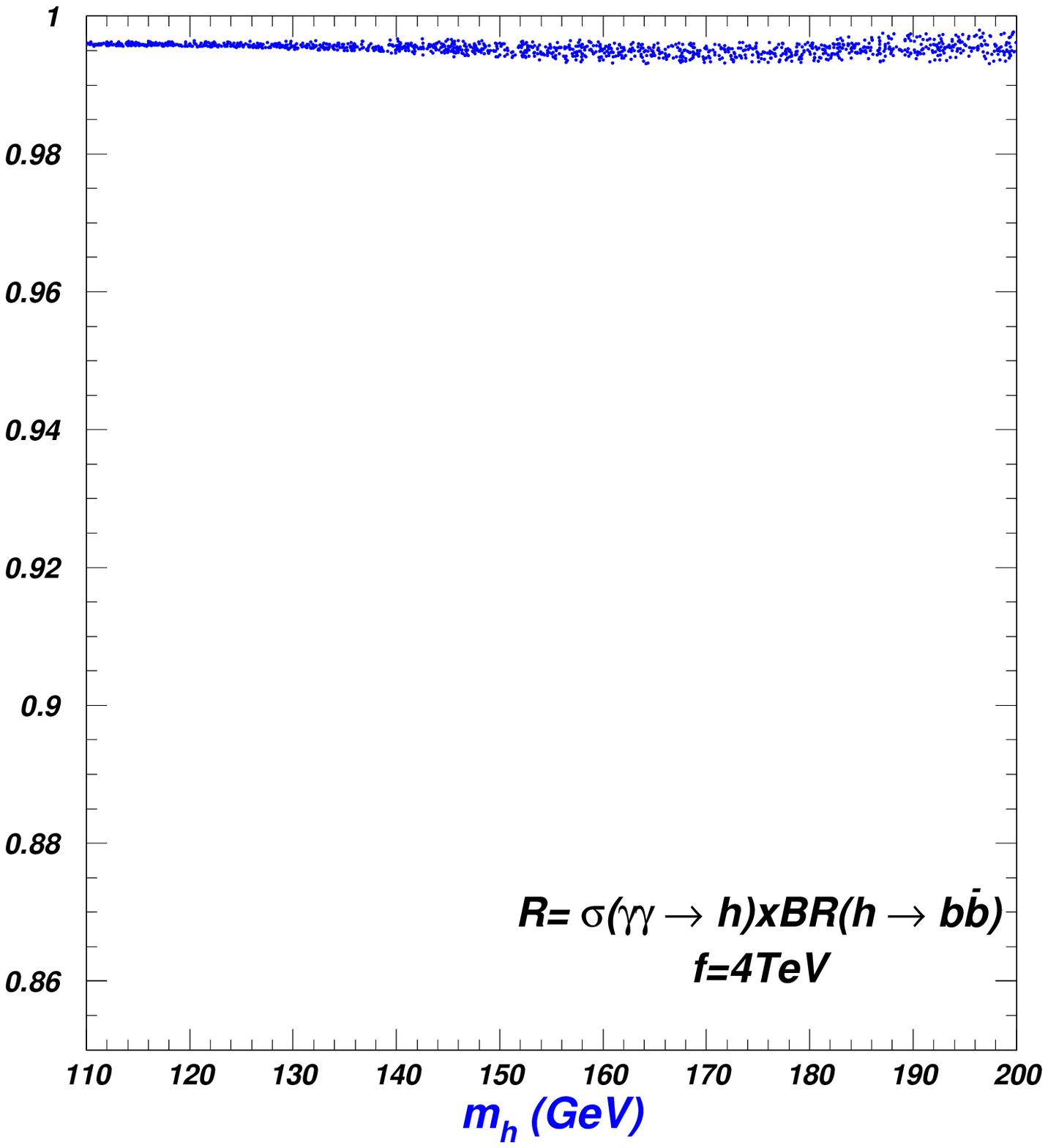,height=5.7cm}
\vspace{-.5cm}
\caption{Scatter plots for the rate
$\sigma(\gamma\gamma\to h)\times BR(h\to b\bar{b})$ normalized to
the SM prediction in the LH model. } \label{figlh}
\end{figure}

\begin{figure}[htb]
 \epsfig{file=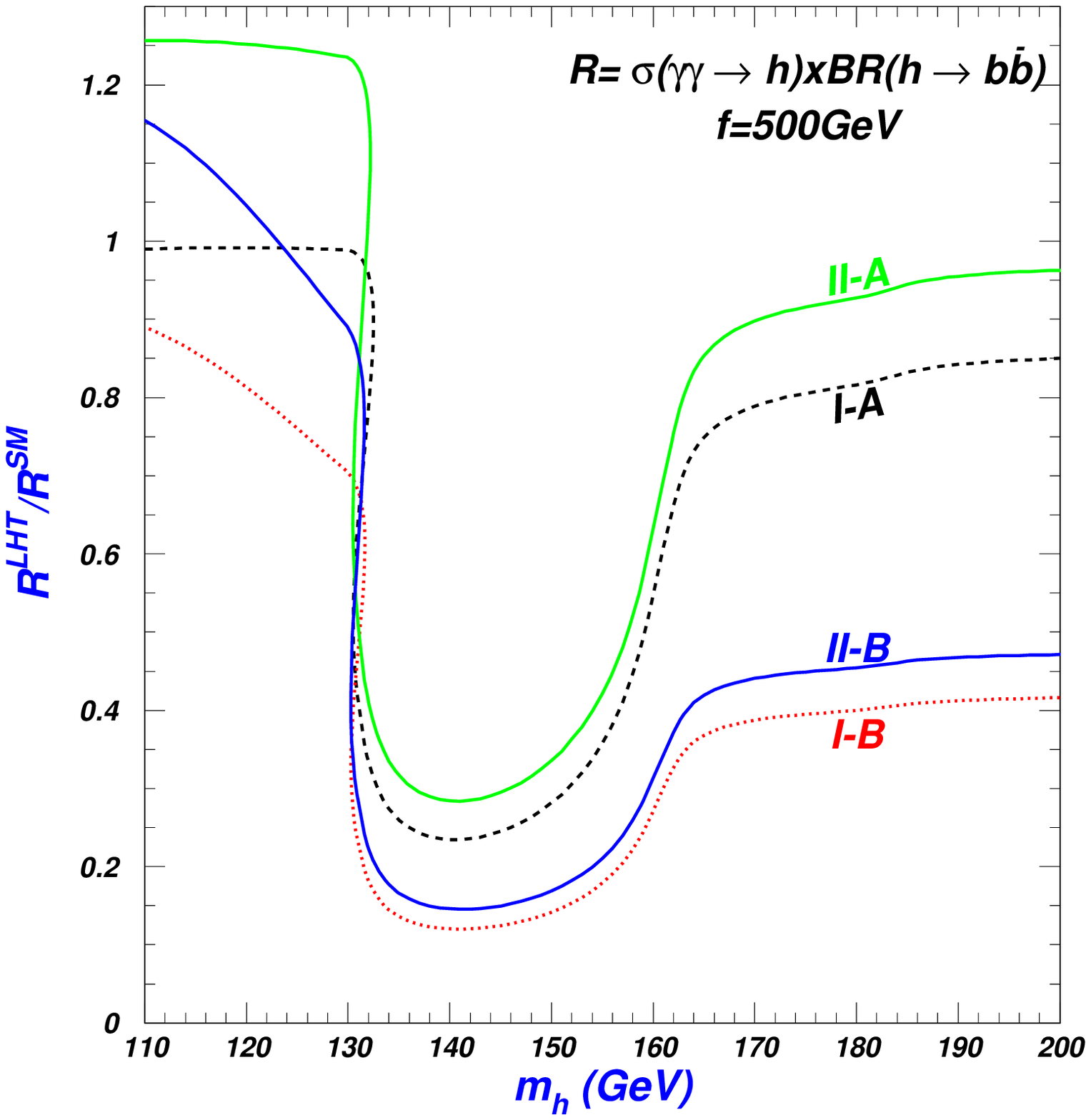,height=5.7cm}
 \epsfig{file=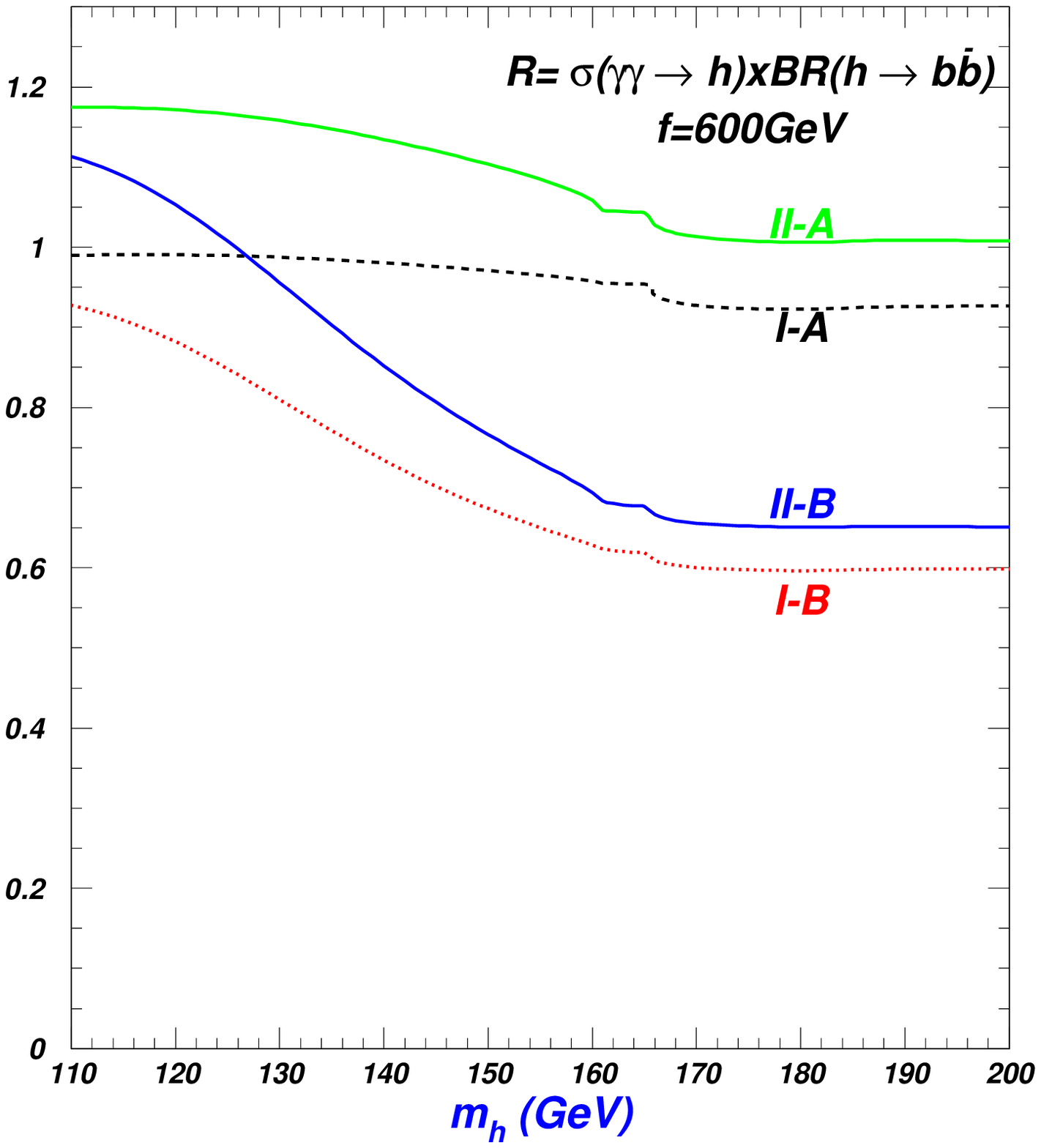,height=5.7cm}
 \epsfig{file=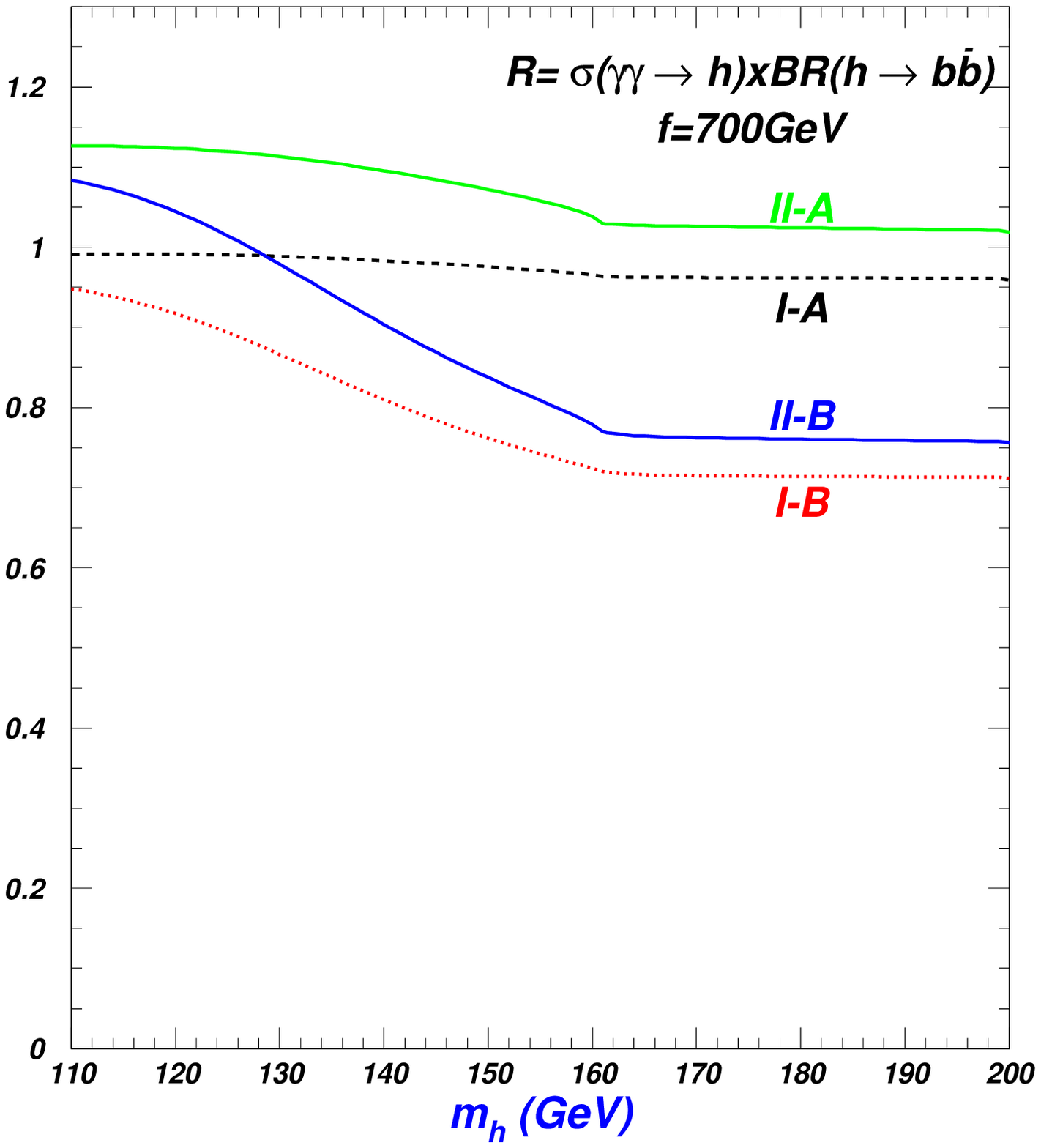,height=5.7cm}
 \epsfig{file=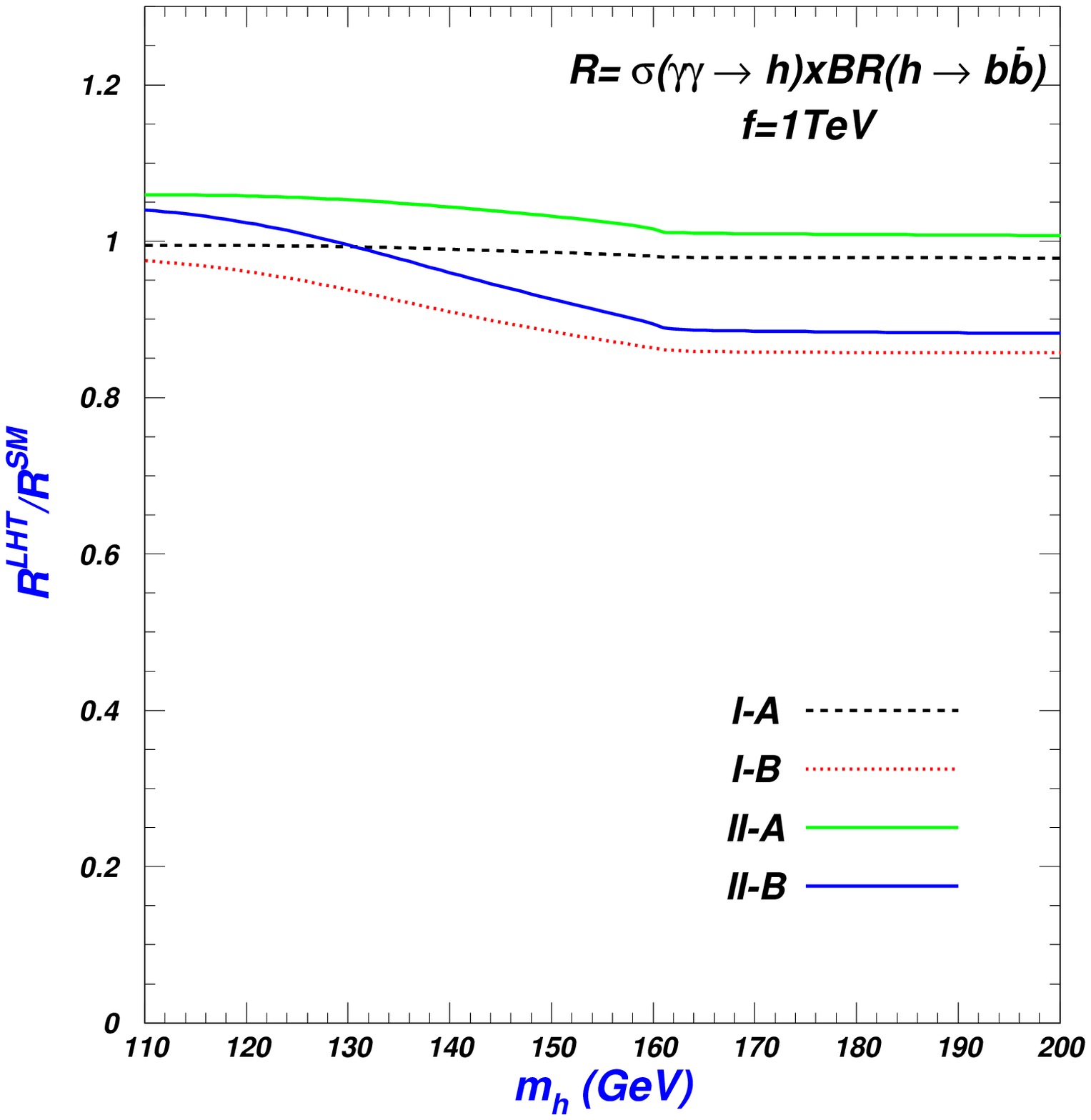,height=5.7cm}
 \epsfig{file=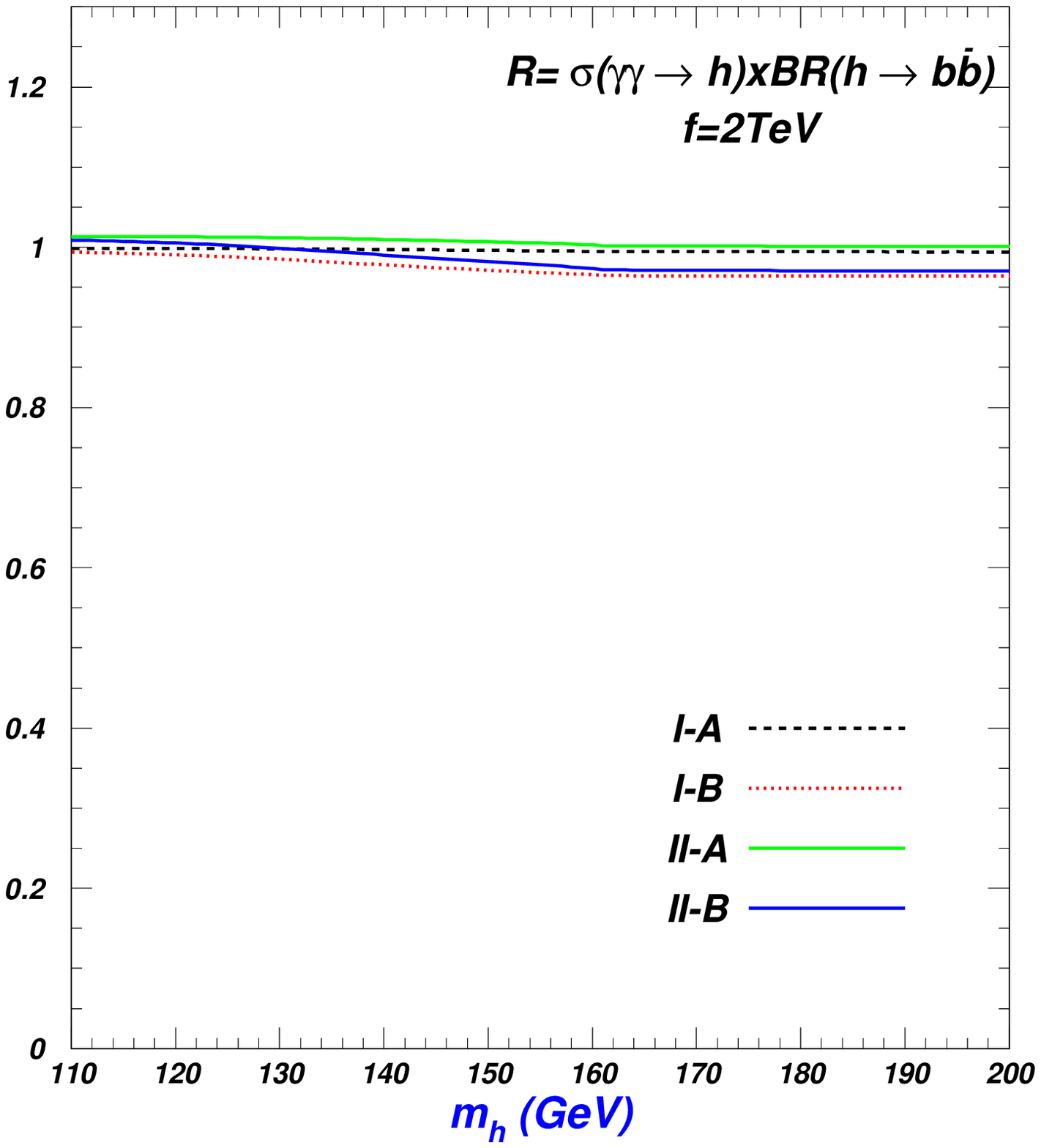,height=5.7cm}
\vspace{-.5cm}
\caption{The rate $\sigma(\gamma\gamma\to
h)\times BR(h\to b\bar{b})$ normalized to the SM prediction in the
LHT-I and LHT-II models. Here I-A (II-A) and I-B (II-B) denote
Case A and Case B, respectively.}
\label{figlht}
\end{figure}

The numerical results for the rate
$\sigma(\gamma\gamma\to h)\times BR(h\to b\bar{b})$
 are shown in Figs. \ref{figslh}, \ref{figlh} and \ref{figlht},
normalized to the SM prediction.
We see that the rate in all these little Higgs models can have a
sizable deviation from the SM prediction, and the magnitude of
deviation is sensitive to the scale $f$.

Fig. \ref{figslh} shows that the SLH model always suppresses the
rate, and the suppression is more sizable for a large $\tan\beta$.
When $\tan\beta$ is large enough, such as $t_\beta=10$ for $f=2$ TeV
($t_\beta=18$ for $f=4$ TeV or $t_\beta=25$ for $f=5.6$ TeV), the
suppression can be as much as $90\%$. The reason for such a severe
suppression is that the decay mode $h\to \eta\eta$ can be dominant
in some part of the parameter space and thus the total decay width
of Higgs boson becomes much larger than the SM value. Note that
$\tan\beta$ cannot be too large for a fixed $f$ in order for the
perturbation to be valid. As shown in Eq. (\ref{eq:v}), the
correction to the Higgs VEV is proportional to $\tan^2\beta
v_{SM}^2/f^2$. If we require $\ord(v_{SM}^4/f^4)/\ord(v_{SM}^2/f^2)
< 0.1$ in the expansion of $v$, the value of $\tan\beta$ should be
below 10, 20, and 28 for $f=2$ TeV, 4 TeV, and 5.6 TeV,
respectively.

Fig. \ref{figlh} shows that the LH model also always suppresses the
rate $\sigma(\gamma\gamma\to h)\times BR(h\to b\bar{b})$, but the
suppression can only reach about $10\%$. For a light Higgs boson or
a large value of $f$, the suppression is small and not sensitive to
the parameters $c$, $c'$, $c_t$ and $x$. For example, for $f=2$ TeV
the suppression is only a few percent.

Fig. \ref{figlht} shows that LHT-I always suppresses the rate but
LHT-II can either suppress or enhance the rate, depending on the
values of the Higgs mass and scale $f$. For each model the rate in
Case A is always above the rate in Case B because the $hb\bar{b}$
coupling in Case A is less suppressed than in Case B. Also, we see
that for $f=500$ GeV and $m_h$ in the range of $130-150$ GeV, the
rate in both models drops drastically. The reason for such a severe
suppression is similar to what happens in the SLH model discussed
above, i.e., the opening of new decay mode (but now the new mode is
$h\to A_H A_H$).

>From our above results we see that the Higgs production process
$\gamma\gamma\to h \to b\bar{b}$ can be a powerful probe for various
little Higgs models. Also, as shown in the literature \cite{cao},
the photon-photon collision option of the ILC can probe the
top-quark related new physics more effectively than in the $e^+e^-$
collision. Therefore, such a photon-photon collision option is well
motivated from the viewpoint of probing new physics.

\section{Conclusion}
We studied the process $\gamma\gamma\to h \to b\bar{b}$ at the
photon-photon collision of the ILC as a probe of different little
Higgs models, including the SLH, LH, LHT-I and LHT-II. We obtained
the following observations: (i) Compared with the SM prediction, the
SLH, LH and LHT-I always suppress the rate of $\gamma\gamma\to h \to
b\bar{b}$; while the LHT-II can either suppress or enhance the rate,
depending on the values of the Higgs mass and scale $f$; (ii) The
deviation of the production rate from its SM prediction is sensitive
to the scale $f$ in all these models. In the SLH, the deviation  is
also sensitive to $\tan\beta$; (iii) The production rates in the SLH
and LHT-I/LHT-II can be severely suppressed in some part of the
parameter space where the new decay mode, $h\to \eta\eta$ for the
SLH and $h\to A_H A_H$ for the LHT-I/LHT-II, is open and dominant.
Therefore, the precision measurement of such a production process at
the ILC will allow for a test of these models and even distinguish
between different scenarios.

\section*{Acknowledgment}
This work was supported in part by the Foundation of Yantai University
under Grant No.WL09B31, by the National Natural Science
Foundation of China (NNSFC) under grant Nos. 10821504,
10725526 and 10635030, by
the Project of Knowledge Innovation Program (PKIP) of Chinese
Academy of Sciences under grant No. KJCX2.YW.W10 and by an
invitation fellowship of LHC Physics Focus Group, National Center
for Theoretical Sciences, Taiwan, Republic of China.

\appendix
\section{The Effective couplings of
              Higgs-photon-photon and Higgs-gluon-gluon}

The effective Higgs-photon-photon coupling can be written as
\cite{hrrhan,higgshunter} \beq {\cal
L}^{eff}_{h\gamma\gamma}=-\frac{\alpha}{8\pi
v}IF_{\mu\nu}F^{\mu\nu}h, \label{hrr} \eeq where $F^{\mu\nu}$ is the
electromagnetic field strength tensor. With the Higgs boson
couplings to the charged fermion $f_i$, vector boson $V_i$ and
scalar $S_i$ given by \beq
 {\cal L}=\sum_{f_i}-\frac{m_{f_i}}{v}y_{_{f_i}}\bar{f_i}f_i h+ \sum_{V_i}2
 \frac{m_{_{V_i}}^2}{v} y_{_{V_i}} V_i V_i h
   +\sum_{S_i}- 2 \frac{m_{_{S_i}}^2}{v} y_{_{S_i}} S_i S_i h,
\label{rrinteri} \eeq the factor $I$ in Eq. (\ref{hrr}) can be
written as \beq I=\sum_{f_i} Q_{f_i}^2 N_{cf_i}~y_{_{f_i}}
I_{\frac{1}{2}}(\tau_{_{f_i}})+\sum_{V_i} Q_{V_i}^2
~y_{_{V_i}}I_1(\tau_{_{V_i}})+\sum_{S_i} Q_{S_i}^2
~y_{_{S_i}}I_0(\tau_{_{S_i}}), \eeq where $Q_X$ ($X$ denotes $f_i$,
$V_i$ and $S_i$) is the electric charge for a particle $X$ running
in the loop, and $N_{cf_i}$ is the color factor for $f_i$. The
dimensionless loop factors are
\begin{eqnarray}
I_{\frac{1}{2}}(\tau_{_{f_i}})& =& -2\tau_{_{f_i}} [1 +
   (1-\tau_{_{f_i}})f(\tau_{_{f_i}})], \\
I_1(\tau_{_{V_i}}) &=& 2 + 3 \tau_{_{V_i}} + 3\tau_{_{V_i}}(2-\tau_{_{V_i}})
    f(\tau_{_{V_i}}),\\
I_0(\tau_{_{S_i}}) &=& \tau_{_{S_i}} [1 - \tau_{_{S_i}} f(\tau_{_{S_i}})],
\end{eqnarray}
where $\tau_{_X} =4m_X^2/m_h^2$ and
\begin{equation}
    f(\tau_{_X}) = \left\{ \begin{array}{lr}
        [\sin^{-1}(1/\sqrt{\tau_{_X}})]^2, & \tau_{_X} \geq 1 \\
        -\frac{1}{4} [\ln(\eta_+/\eta_-) - i \pi]^2, & \, \tau_{_X} <
        1
        \end{array}  \right.\label{hggf12}
\end{equation}
with $\eta_{\pm}=1\pm\sqrt{1-\tau_X}$.  When the masses of particles
in the loops are much larger than half of the Higgs boson mass, we
can get
\begin{equation}
I_{\frac{1}{2}}(\tau_{_{f_i}}) \simeq  -4/3, \qquad
I_1(\tau_{_{V_i}}) \simeq  7, \qquad
I_0(\tau_{_{S_i}}) \simeq -1/3.
\end{equation}
The effective Higgs-gluon-gluon coupling can be written as
\cite{hrrhan,higgshunter} \beq {\cal
L}^{eff}_{hgg}=-\frac{\alpha_s}{12\pi v}I_{hgg}
G_{\mu\nu}^{\alpha}G^{\mu\nu}_{\alpha}h, \label{hgg} \eeq where
$G_{\mu\nu}^{\alpha}=\partial_{\mu} g_{\nu}^{\alpha}-\partial_{\nu}
g_{\mu}^{\alpha}$ and the factor $I_{hgg}$ from the contributions of
quarks running in the loops is given by \beq I_{hgg}=\sum_{q_i}
\frac{3}{4}y_{_{q_i}} I_{\frac{1}{2}}(\tau_{_{q_i}}), \eeq with
$\tau_{_{q_i}}=4m_{q_i}^2/m_h^2$.

Once the interactions in Eq. (\ref{rrinteri}) are given, we can
obtain the effective $h\gamma\gamma$ and $hgg$ couplings from the
above formulas. In the following we list the relevant Higgs
interactions in the SLH, LH, LHT-I and LHT-II, respectively. Here
the Higgs interactions with the light fermions are not listed since
their contributions can be ignored.
\begin{itemize}
\item[(1)] In the SLH, the Higgs couplings with the quarks are given by
\bea \label{tTmixing}
{\cal L}_t &\simeq&-f \lambda_2^t \left[ x_\lambda^t c_\beta
t_1^{c'}(-s_1t'_L
   +c_1T'_L)+s_\beta t_2^{c'} (s_2 t'_L+ c_2 T'_L)\right]+h.c.,\,\\
   \label{dDmixing}
{\cal L}_{d} &\simeq&-f \lambda_2^{d} \left[ x_\lambda^{d} c_\beta
d_1^{c'}
  (s_1 d'_{L}+c_1 D'_{L})+s_\beta d_2^{c'} (-s_2 d'_{L}+c_2 D'_{L})\right]+h.c.,\,\\
\label{sDmixing} {\cal L}_{s} &\simeq&-f \lambda_2^{s} \left[
x_\lambda^{s} c_\beta s_1^{c'}
  (s_1 s'_{L}+c_1 S'_{L})+s_\beta s_2^{c'} (-s_2 s'_{L}+c_2 S'_{L})\right]+h.c.,\, ,
\eea
where
\bea
s_1\equiv \sin {t_\beta (h+v)\over \sqrt{2}f},\ \
s_2\equiv \sin{(h+v) \over \sqrt{2}t_\beta f},\ \ s_3\equiv
\sin{(h+v)(t_\beta^2+1)\over \sqrt{2}t_\beta f}.
\eea
After diagonalization of the mass matrix in Eqs. (\ref{tTmixing}),
(\ref{dDmixing}) and (\ref{sDmixing}), we can get the mass
eigenstates $(t,~T)$, $(d,~D)$ and $(s,~S)$, which was performed
numerically in our analysis, and the relevant couplings with Higgs
boson can be obtained without resort to any expansion of $v/f$ (
the diagonalization of the quark mass matrix in the LH, LHT-I and
LHT-II was also performed numerically in our calculation).

The Higgs coupling with the bosons is given by \cite{slhvdefine},
\beq
 {\cal L}=
 2\frac{m_{W}^2}{v} y_{_W} W^+ W^- h+2\frac{m_{W'}^2}{v} y_{_{W'}} W^{'+} W^{'-} h,\eeq
where \beq y_{_W} \simeq \frac{v}{v_{SM}} \left[ 1- \frac{v_{SM}^2}{
4f^2}\frac{\tbt^4-\tbt^2+1}{\tbt^2} + \frac{v_{SM}^4}{36
f^4}\frac{(\tbt^2-1)^2}{\tbt^2} \right],\qquad y_{_{W'}} \simeq
-\frac{v^2}{2f^2}.
\eeq

\item[(2)] In the LH, the Higgs couplings with the heavy quarks are given by
\beq {\cal L}_t\simeq -\lambda_1 f \left[\frac{s_\Sigma}{\sqrt{2}}
\bar{u}_{L}u_R+ \frac{1+c_\Sigma}{2} \bar{U}_{L}u_R
\right]-\lambda_2f\bar{U}_LU_R+{\rm h.c.}, \label{lhtop} \eeq where
$c_\Sigma\equiv \cos\frac{\sqrt{2}(v+h)}{f}$ and $s_\Sigma\equiv
\sin\frac{\sqrt{2}(v+h)}{f}$. After diagonalization of the mass
matrix in Eq. (\ref{lhtop}), we can get the mass eigenstates $t$ and
$T$ as well as their couplings with the Higgs boson \cite{hrrhan}:
\beq
 {\cal L}=
 -\frac{m_t}{v} y_{t} \bar{t}t h - \frac{m_T}{v} y_{_T}
 \bar{T}T h,
\eeq
where
\beq
y_{t} = 1+\frac{v^2}{f^2} \left[ -
\frac{2}{3}+\frac{x}{2}-\frac{x^2}{4} + c_t^2 s_t^2 \right],\qquad
y_{_{T}} = -c_t^2s_t^2\frac{v^2}{f^2}.
\eeq
The Higgs coupling with the bosons are given by
\begin{eqnarray}
 {\cal L}&=&
 2\frac{m_{W}^2}{v} y_{_W} W^+ W^- h
 + 2\frac{m_{W_H}^2}{v} y_{_{W_H}} W^{+}_H W^{-}_H h \nonumber\\
&&
 -2\frac{m_{\Phi}^2}{v} y_{_{\Phi^+}} \Phi^+ \Phi^- h
 - 2\frac{m_{\Phi}^2}{v} y_{_{\Phi^{++}}} \Phi^{++}
  \Phi^{--} h,
\end{eqnarray}
where
\begin{eqnarray}
    \label{yi}
    \begin{array}{rclrcl}
    y_{_{W_L}}&=&1+ \frac{v^2}{f^2}\left[-\frac{1}{6}-\frac{1}{4}(c^2-s^2)^2\right],
                   \qquad &
    y_{_{W_H}}&=&- s^2c^2\frac{v^2}{f^2}, \\
    y_{\Phi^+}&=& \frac{v^2}{f^2}\left[-\frac{1}{3}+\frac{1}{4}x^2\right], \qquad &
    y_{\Phi^{++}}&=& \frac{v^2}{f^2}\ord(\frac{x^2}{16}\frac{v^2}{f^2},\frac{1}{16\pi^2}).
    \end{array}
\end{eqnarray}
Since the $h\Phi^{++}\Phi^{--}$ coupling is very small, the
 contributions of the doubly-charged scalar can be ignored.

\item[(3)] In the LHT-I, the Higgs couplings with the heavy quarks are given by
\begin{eqnarray}
{\cal L}_{\kappa}&\simeq& -\sqrt{2} \kappa f
\left[\frac{1+c_\xi}{2} \bar{u}_{L_-} u'_R
-\frac{1-c_\xi}{2}\bar{u}_{L_-}q_R
-\frac{s_\xi}{\sqrt{2}}\bar{u}_{L_-} \chi_R\right] \nonumber \\
&& - m_q
\bar{q}_Lq_R - m_{\chi}\bar{\chi}_L\chi_R +{\rm
h.c.},\label{lhti-odd}
\end{eqnarray}
\beq {\cal L}_t\simeq -\lambda_1 f \left[\frac{s_\Sigma}{\sqrt{2}}
\bar{u}_{L_+}u_R+ \frac{1+c_\Sigma}{2} \bar{U}_{L_+}u_R
\right]-\lambda_2f \bar{U}_{L_+}U_{R_+}+{\rm h.c.}, \label{lhti-t}
\eeq where $c_\xi \equiv \cos\frac{v+h}{\sqrt{2}f}$ and $s_\xi\equiv
\sin\frac{v+h}{\sqrt{2}f}$. After diagonalization of the mass matrix
in Eq. (\ref{lhti-odd}), we can get the T-odd mass eigenstates
$u_-$, $q$ and $\chi$. In fact, there are three generations of T-odd
particles, and we assume they are degenerate. The mass eigenstates
$t$ and $T$ can be obtained by mixing the interaction eigenstates in
Eq. (\ref{lhti-t}).

The Higgs interactions with the bosons in the LHT-I can be obtained
from the couplings in the LH by taking $c=s=1/\sqrt{2}$ and $x=0$.

\item[(4)] In the LHT-II,
the Higgs couplings with the first two generations of heavy quarks
are given by
\beq
\label{lhtii-12} {\cal L}_{q}^{1,2}\simeq-\sqrt{2} \kappa f
\left[\frac{1+c_\xi}{2} \bar{u}_{L_-} u'_R
-\frac{1-c_\xi}{2}\bar{u}_{L_-}q_R
+\frac{s_\xi}{\sqrt{2}}\bar{u}_{L_+} \chi_R \right]- m_q
\bar{q}_Lq_R- m_{\chi}\bar{\chi}_L\chi_R+{\rm h.c.} .
\eeq
The mass eigenstates of $u_-$, $q$ and $\chi$ can be obtained by
the diagonalization of the mass matrix in Eq. (\ref{lhtii-12}).

The Higgs couplings with the third generation of heavy quarks are given by
\begin{eqnarray} {\cal L}_{q}^{3}
&\simeq&-\sqrt{2} \kappa f \left [\frac{1+c_\xi}{2} \bar{u}_{L_-}
u'_R -\frac{1-c_\xi}{2}\bar{u}_{L_-}q_R
-\frac{s_\xi}{\sqrt{2}}\bar{U}_{L_-}q_R
  -\frac{s_\xi}{\sqrt{2}}\bar{U}_{L_-}u'_R
  +\frac{s_\xi}{\sqrt{2}}\bar{u}_{L_+} \chi_{R}
  \right. \nonumber \\&&\left.+c_\xi\bar{\chi}_{L}\chi_{R} \right]- m_q
\bar{q}_Lq_R -\lambda f \left[s_\Sigma
\bar{u}_{L_+}u_{R_+}+\frac{1+c_\Sigma}{\sqrt{2}} \bar{U}_{L_-}
U_{R_-} \right]+{\rm h.c.}, \label{lhtii-3}
\end{eqnarray}
where $c_t$ is taken as $1/\sqrt{2}$. After diagonalization of the
mass matrix in Eq. (\ref{lhtii-3}), we can get the mass eigenstates
$t$, $T_-$, $u_-$, $q$ and $\chi$.

The Higgs interactions with the bosons in the LHT-II are the same as
in the LHT-I.
\end{itemize}

Note that in the lepton sector, the SLH, LHT-I and LHT-II also predict some
neutral heavy neutrinos, which do not contribute to the couplings of
$h\gamma\gamma$ and $hgg$ at the one-loop level.
Although the charged heavy leptons and down-type T-odd quarks are
predicted in LHT-I and LHT-II, they do not have direct couplings with
the Higgs boson.
Besides, from Eqs. (\ref{hrr}) and (\ref{hgg}), we can find that the
effective couplings of $h\gamma\gamma$ and $hgg$ are related to the
Higgs VEV $v$ and the running $\alpha$ and $\alpha_s$ in these models.

\end{document}